\documentstyle[latexsym,amssymb,amsfonts]{article}
\def\ket#1{\left|#1\right>}

\def\bracket#1#2{\left<#1\right.\left|#2\right>}
\newcommand{\nad}[1]{\mbox{\smash{\oalign{$#1$ \crcr \hidewidth 
$\mathchar"017E$ \hidewidth}}}}
\newcommand{\Nad}[1]{\mbox{\smash{\oalign{$#1$ \crcr \hidewidth 
$\mathchar"707E$ \hidewidth}}}}
\begin{document}
\begin{center}
{\LARGE\bf Lorentz-covariant quantum mechanics\\[2mm]
and preferred frame}
\end{center}
\vspace{1cm}
\begin{center}
{\large P.\ Caban\footnote{E-mail address: caban{@}mvii.uni.lodz.pl}}
{\large and  J.\ Rembieli\'nski\footnote{E-mail address:
jaremb{@}mvii.uni.lodz.pl and jaremb{@}krysia.uni.lodz.pl}}
\end{center}
\begin{center}
Department of Theoretical Physics, University of {\L}\'od\'z,\\
Pomorska 149/153, 90-236 {\L}\'od\'z, Poland
\end{center}
\vspace{.5cm}
\begin{center}
\the\year--\the\month--\the\day
\end{center}
\vspace{1cm}
\begin{center}
{\bf Abstract}
\end{center}
\vspace{2mm}
\begin{center}
\begin{tabular}{c}
\newlength{\str}
\setlength{\str}{\textwidth}
\addtolength{\str}{-2cm}
\parbox{\str}{\small
In this paper the relativistic quantum mechanics is considered
in the framework of the nonstandard synchronization scheme for
clocks. Such a synchronization
preserves Poincar{\'e} covariance but (at least formally) distinguishes
an inertial frame. This enables to avoid the problem of a noncausal
transmision of information related to breaking of the Bell's
inequalities in QM. Our analysis has been focused
mainly on the problem of existence of a proper position operator
for massive particles. We have proved that in our framework such an
operator 
exists for particles with arbitrary spin. It fulfills all the
requirements: it is Hermitean and covariant, it has commuting
components and moreover its eigenvectors (localised states) are
also covariant. We have found the explicit form of the position
operator and have demonstrated that
in the preferred frame our operator coincides with the
Newton--Wigner one. We have also defined a covariant spin operator and
have constructed an invariant spin square operator. Moreover, full
algebra of observables consisting of position operators, fourmomentum
operators and spin operators is manifestly Poincar\'e covariant in
this framework. Our results support expectations of other authors 
(Bell \cite{Bell81}, Eberhard \cite{Eberhard78}) that a
consistent formulation of quantum mechanics demands existence of a
preferred frame.} 
\end{tabular}
\end{center}
\vspace{4mm}

\section{Introduction}
In this paper we propose a formulation of the Poincar\'e covariant
quantum mechanics for a free particle. Our investigations were
motivated by two old and still open problems: violation of locality
in quantum mechanics (breaking of Bell's inequalities) and
nonexistence of a covariant position operator as well as covariant
localized states. 
It was recognised long time ago, that some correlation experiments
(like Aspect {\em et al} \cite{Aspect82a,Aspect81,Aspect82b} ones)
imply, that, as was stressed by H.\ P.\ 
Stapp \cite{Stapp77} ``... what happens macroscopically in one space-time
region must in some cases depend on variables that are controlled by
experimenters in far-away, space-like-separated regions''. This fact
can be in a conflict with special relativity; even more frustrating is
a conflict 
with a causal transmission of information. 
It may be interesting to recall in this
place the statement by J.\ S.\ Bell \cite{Bell86}: ``... For me then
this is the 
real problem with quantum theory: the apparently essential conflict
between any sharp formulation and fundamental relativity''.
According to Bell \cite{Bell81} (see also Eberhard \cite{Eberhard78})
consistent formulation of relativistic quantum mechanics may be
necessarily with a preferred frame at the fundamental level.
Following these suggestions we try to construct here a quantum
mechanics which has built--in the preferred frame and which is at the
same time Poincar\'e covariant. It is important to realise at this
point, that special relativity is in fact based on two
main assumptions: Poincar\'e covariance and relativity
principle. 
So the key point is to reformulate this theory in
a way which preserves the Poincar\'e covariance but abandons the
relativity principle and consequently allows us to introduce a
preferred frame. Such a formulation of relativity theory has been
given by one of the authors (J.R.) in 
\cite{Remb80,Remb1}. It was shown
there, that, using a nonstandard synchronization procedure for
clocks (named in \cite{Remb1} as the
Chang--Tangherlini synchronization), it is possible to obtain
such a form of the transformations of coordinates beetwen
inertial observers, that they realize Lorentz transformations, the
time coordinate
is only rescaled by a positive factor and the space coordinates
do not mix to it.  A price for this is existence of a preferred
frame in the theory  
and dependence of Lorentz group transformations on the
additional parameter --- the fourvelocity of 
the preferred frame.
Usually it is claimed that existence of a preferred frame
violates the Poincar\'e covariance. This is really the case when we
restrict ourselves to the Minkowski space-time. But in our approach
the additional set of parameters (the mentioned above fourvelocity)
allows us to solve this difficulty.
That is this theory preserves Poincar\'e covariance but not
necessarily the
relativity principle. However, in our opinion, this is not any serious
problem: in the real (expanding) Universe such a frame really does
exist --- it is the so called comoving frame related to the matter
and the cosmic background radiation frame.
Furthermore, in our framework, average light velocity over closed
paths is still constant 
and equals to $c$, so Michelson-Morley like experiments do not
distinguish such a possibility from the standard one
\cite{Jammer79,Remb1}. 
Moreover, in the context of the Poincar\'e 
covariant quantum mechanics presented herein,
the above mentioned conflict  between 
causality and quantum theory disappears. 
We also hope, that this formulation will be very convenient in the
case of 
Poincar\'e covariant de Broglie--Bohm approach to 
quantum mechanics\footnote{An eshaustive review of the
  interpretational problems of quantum mechanics can be found in
\cite{Bell88,Holland,Hughes89,Isham,Peres93}.}. 
Next problem which is solved in
this context is the localisation problem.
Various aspects of localizability of particles have been studied
from the early days of quantum mechanics, but, in the
relativistic case (in contradiction to the non-relativistic one)
the fully satisfactory position operator has not been found up to
now. Let us explain at this point what we mean by the satisfactory
position operator. Such an operator should be Hermitean, have
commuting components (for massive particles), fulfill the canonical
commutation relations with the momentum operators, be covariant and 
have covariant eigenstates (localized states). 
Operator
constructed in the framework of the presented here theory fulfills all
of the stated above conditions.
To make this paper self-contained we devote the second section
to the description of the Chang--Tangherlini synchronization scheme
and the corresponding Lorentz group realization.
More details on this subject one can find in \cite{Remb1}.
In the section 3, following \cite{Remb1,RemCab}, we describe
briefly the covariant canonical formalism for a relativistic
free particle on the classical level.
The section 4 contains
description of the quantum theory in the Chang--Tangherlini
synchronization and the definition of the position operator.
In the section 5 we construct and classify unitary orbits of
the Poincar\'e group using the introduced position operator. We define
also covariant spin operator and 
construct the invariant operator of the spin square.
In the section 6 we
firstly give a very brief review
of the properties of the Newton--Wigner operator (in the authors'
opinion the best position operator which has been constructed up
to now; more information about the history of the localization
problem and the full bibliography can be found in \cite{Bac,Nied}).
Then we find the explicit form of our position operator in the
functional realization and show that in the preferred frame our
operator coincides with the Newton-Wigner one. This section
contains also a discussion of the position operator
under the special choice of the integral measure. In this case
the form of the position operator is the same as in the
nonrelativistic quantum mechanics.

\section{The Chang--Tangherlini (CT) synchronization}
In this chapter we briefly describe main results connected
with the CT synchronization scheme which we shall use in the
following. Derivation of these results one can find in \cite{Remb1}.
The idea applied there is based on well known facts that 
the definition of the time coordinate
depends on the choice of the synchronization scheme for clocks
and that this choice is a convention
\cite{Jammer79,Mansouri77,Reichenbach69,Will92,Will93}.  
Using this freedom of choice
one can try to find a synchronization procedure 
resulting in the desired form of the Lorentz transformations.
Performing such a program we have to distinguish, at least formally,
one inertial frame --- so called preferred frame. Thus, at least
formally, the relativity principle may be broken. We discuss this
question in the Section \ref{bundle}. 
Now, each inertial frame is determined by
the fourvelocity of this frame with respect to the
distinguished one. We denote fourvelocity
of the preferred frame as seen by an inertial observer by $u^\mu$.
Hereafter, quantities in the Einstein-Poincar\'e (EP) synchronization
are denoted by the subscript (or superscript) $E$. Quantities
in the CT synchronization will have no index. We use
the natural units ($\hbar=c=1$).
According to \cite{Remb80,Remb1} transformation law beetwen inertial
frames is determined by the following requirements:
\begin{list}{}{
\setlength{\parsep}{0pt}
\setlength{\topsep}{0pt}
\setlength{\itemsep}{0pt}}
\item[$1^0$] the transformation group is isomorphic to the Lorentz
  group;
\item[$2^0$] the average value of the light speed over closed paths
  is constant and equal to 1;
\item[$3^0$] transformations are linear with respect to the
  coordinates (affinity);
\item[$4^0$] under rotations coordinates transform in a standard way
  (isotropy) 
   \[ x^{\prime 0}(u^\prime)=x^0(u),  \]
   \[ \vec{x}^{\prime}(u^\prime)=R\vec{x}(u), \]
  here $R$ is the rotation matrix;
\item[$5^0$] the instant time hyperplane $x^0=const$ is an invariant
  notion.
\end{list}
Notice that $1^0-4^0$ are the standard requirements, while
nonstandard is the last one ($5^0$).
Consequently the transformation of coordinates between inertial frames
in this 
synchronization has the following form (for contravariant
coordinates)
 \begin{equation}
 x^{\prime}(u^{\prime})=D(\Lambda,u)x(u),
 \label{3}
 \end{equation}
where $D(\Lambda,u)$ is a $\Lambda$ and $u$ dependent $4\times4$
matrix, $\Lambda$ -- element 
of the Lorentz group and $u^\mu$ is the fourvelocity of the
preferred frame with respect to considered frame, so (\ref{3}) is
accompanied by
 \begin{equation}
 u^\prime=D(\Lambda,u)u.
 \label{4}
 \end{equation}
Matrices $D(\Lambda,u)$ fulfill the following group composition rule
 \begin{equation}
 D(\Lambda_2,D(\Lambda_1,u)u)D(\Lambda_1,u)=D(\Lambda_2\Lambda_1,u)
 \label{5}
 \end{equation}
so
 \begin{equation}
 D^{-1}(\Lambda,u)=D(\Lambda^{-1},D(\Lambda,u)u),\qquad D(I,u)=I.
 \label{6}
 \end{equation}
Let $T(u)$ be the intertwining matrix connecting coordinates
in CT and EP synchronizations. It means that for every contravariant
fourvector $A^\mu$ we have
 \begin{equation}
 A^\mu=T(u)^{\mu}_{\phantom{\mu}\nu}A^{\nu}_{E}.
 \label{7}
 \end{equation}
Therefore $D(\Lambda,u)$ is of the following form
 \begin{equation}
 D(\Lambda,u)=T(u^\prime)\Lambda T^{-1}(u).
 \label{8}
 \end{equation}
One can find the explicit form of $T(u)$ (\cite{Remb1}),
namely
 \begin{equation}
 T(u)=\left(\begin{array}{c|c}
 1 & -\vec{u}^T u^0\\
 \hline
 0 & I
 \end{array}\right).
 \label{9}
 \end{equation}
Consequently $D(\Lambda,u)$ is given for rotations ($R\in SO(3)$) by
 \begin{equation}
 D(R,u)=\left(\begin{array}{c|c}
 1 & 0 \\
 \hline
 0 & R
 \end{array}\right);
 \label{10}
 \end{equation}
while for boosts by
 \begin{equation}
 D(W,u)=\left(\begin{array}{c|c}
 \frac{1}{W^0} & 0 \\
 \hline
 -\vec{W} & I + \frac{\vec{W}\otimes\vec{W}^T}%
 {\left(1+\sqrt{1+(\vec{W})^2}\right)} - u^0\vec{W}\otimes\vec{u}^T
 \end{array}\right),
 \label{11}
 \end{equation}
where $W^\mu$ denotes the fourvelocity of the primed frame
$O_{u^\prime}$ with respect to the frame $O_u$.
Fourvelocity $W^\mu$ can be expressed by $u$ and $u^\prime$
 \begin{equation}
 W^0=\frac{u^0}{u^{\prime 0}},\quad
 \vec{W}=\frac{(u^0+u^{\prime 0})(\vec{u}-\vec{u}^\prime)}%
 {\left[1+u^0u^{\prime 0}(1+\vec{u}\vec{u}^\prime)\right]}.
 \label{12}
 \end{equation}
Instead of $W^\mu$ we can use also velocity
$\vec{V}=\frac{\vec{W}}{W^0}$. The corresponding form of all given
above formulas in such a parametrization can be found by means of
 \begin{equation}
 \frac{1}{W^0}=\sqrt{\left(1+u^0 \vec{u} \vec{V}\right)^2-
 \left(\vec{V}\right)^2}.
 \label{13}
 \end{equation}
The explicit relationship beetwen coordinates in EP and CT is given by
 \begin{equation}
 \begin{array}{ll}
 x_{E}^{0}=x^0+u^0\vec{u}\vec{x},\qquad  & \vec{x}_E=\vec{x},\\
 u_{E}^{0}=\frac{1}{u^0}, & \vec{u}_E=\vec{u}.
 \end{array}
 \label{14}
 \end{equation}
We see that only the time coordinate changes.
Note also, that in the same space point we have $\Delta
x_{E}^{0}=\Delta x^0$ 
so the time lapse is the same in both synchronizations.\\
One can easily see that the line element
 \begin{equation}
 ds^2=g_{\mu\nu}(u) dx^\mu dx^\nu,
 \label{15}
 \end{equation}
is invariant under the transformations (\ref{8}) if
 \begin{equation}
 g(u)=\left(T(u) \eta T^T(u)\right)^{-1},
 \label{16}
 \end{equation}
where $\eta$ is the Minkowski metric tensor
$\eta={\rm diag}(+,-,-,-)$.
The explicit form of the covariant metric tensor reads
 \begin{equation}
 \left[g_{\mu\nu}\right]=
 \left(\begin{array}{c|c}
 1 & u^0 \vec{u}^T \\
 \hline
 u^0 \vec{u} & -I + (u^0)^2 \vec{u}\otimes\vec{u}^T
 \end{array}\right),
 \label{17}
 \end{equation}
while the contravariant one has the form
 \begin{equation}
 g^{-1}(u)=
 \left(\begin{array}{c|c}
 (u^0)^2 & u^0 \vec{u}^T \\
 \hline
 u^0 \vec{u} & -I 
 \end{array}\right),
 \label{18}
 \end{equation}
so the space line element is the Euclidean one: $dl^2=d\vec{x}^2$.
Let us notice here that the triangular form of the boost matrix
(\ref{11}) 
implies, that under the Lorentz transformations the time coordinate
is only rescaled by a positive factor
$(x^{\prime 0}=\frac{1}{W^0}x^0)$; the space coordinates do not
mix to it.
One can also easily check that the following, very useful,
relations hold
 \begin{equation}
 \nad{u}=\nad{0}\qquad{\rm and}\qquad
 \frac{1}{(u^0)^2} - (\vec{u})^2 = 1.
 \label{19}
 \end{equation}
Heareafter the three vector
part of a covariant (contravariant) fourvector $a_\mu$ ($a^\mu$) will
be denoted by $\nad{a}$ ($\vec{a}$) respectively.

\subsection{Geometric description of the CT synchronization}
\label{bundle} 
We can also provide a geometric description of the special
relativity in the CT synchronization scheme in the language of 
frame bundles. To do this, let us denote:\\
$M$ -- the Minkowski space--time,\\
$L_{+}^{\uparrow}$ -- ortochronous Lorentz group 
(the group of space--time transformations);\\
$F(L_{+}^{\uparrow})$ -- the set of all frames in the space $M$ 
obtained by action of $L_{+}^{\uparrow}$ on one particular (but arbitrary)
frame in the space $M$; thus $F(L_{+}^{\uparrow})$ is isomorphic to
the group $L_{+}^{\uparrow}$. 
An element of $F(L_{+}^{\uparrow})$ corresponding, by means of this
isomorphism, to the  
element $g\in L_{+}^{\uparrow}$ is designated by $e(g)$.\\
Now let us consider the following structure
 \begin{equation}
 M_w=\left[L_{+}^{\uparrow},(F(L_{+}^{\uparrow})
 \times M,M,pr_2),\pi_w,\psi_w \right],
 \end{equation} 
where 
$pr_2$ is the canonical projection on the second factor of the
cartesian product; therefore $(F(L_{+}^{\uparrow})\times M,M,pr_2)$ 
is a frame bundle with the typical fibre $F(L_{+}^{\uparrow})$.
$\pi_w$ is a projection on a fixed time-like fourvector $w$, while
$\psi_w$ is the action of the group $L_{+}^{\uparrow}$ on the bundle
$(F(L_{+}^{\uparrow})\times M,M,pr_2)$ fulfilling the following
conditions:\\ 
 \begin{equation}
 (e^\prime(g),x^\prime)=(e(kg),x),
 \label{cond1}
 \end{equation}
 \begin{equation}
 e^{\mu}(kg,x)={D(k,g)^\mu}_\nu e^\nu(g,x),
 \label{cond2}
 \end{equation}
 \begin{equation}
 D^{T-1}(k,g)\pi_w D^{-1}(k,g)=\pi_w.
 \label{cond3}
 \end{equation}
where $k\in L_{+}^{\uparrow}$, $x\in M$. It is clear that we
consider here Lorentz 
transformations as passive transformations --- the action of the
Lorentz group changes the observer, not the physical state. In our
language it means that the action of the Lorentz group changes the
frame $e(g)$. 
The condition (\ref{cond1}) means that the action $\psi_w$ is trivial
on the manifold $M$; the group $L_{+}^{\uparrow}$ acts only in the
fiber. The second  
condition (\ref{cond2}) says that the action $\psi_w$ is linear on
frames.
Now, we associate the time direction with $w$
which means that the projector $\pi_w=\frac{w\otimes w}{w^2}$ is equal
to $\pi_{e^0}$
 \begin{equation}
 \pi_{e^0}=\pi_w,  \label{cond4}
 \end{equation}
so, after this identification, $\pi_w$ in the $e^\mu$ basis
($\pi_w=(\pi_w)_{\mu\nu}e^\mu\otimes e^\nu$) reads
 \begin{equation}
 \left[\left(\pi_{e^0}\right)_{\mu\nu}\right]=
 \left[\left(\pi_{w}\right)_{\mu\nu}\right]=\left(\begin{array}{cccc}
 1 & 0 & 0 & 0\\
 0 & 0 & 0 & 0\\
 0 & 0 & 0 & 0\\
 0 & 0 & 0 & 0
 \end{array}\right).
 \label{cond5}
 \end{equation}
This construction defines a time-orientation of $M$ along $w$.
Now the matrix $D(k,g)$ can be expressed by $D(\Lambda,u)$ given in
eqs.\ (\ref{10}--\ref{11}) as follows: let $\Lambda_1=k$,
$\Lambda_2=g$, then
 \begin{equation}
 D(k,g)=D(\Lambda_1,\Lambda_2\tilde{u}),
 \label{cond6}
 \end{equation}
and $\tilde{u}=(1,\vec{0})$.
The last condition (\ref{cond3}) means that the direction of
the fourvector $w$ is invariant under the action of the group
$L_{+}^{\uparrow}$.\\ 
Thus we have a collection of time-oriented space-times $M_w$, where
$w$ is the 
arbitrary time-like fourvector. The objects $M_w$ and
$M_{w^\prime}$ corresponding to different $w$ and $w^\prime$ are
evidently connected by the action of another Lorentz group
$L_{+}^{\uparrow\,(S)}$ (the 
so called synchronization group --- see \cite{Remb1}). 
The whole family of time-oriented space-times $M_w$
together with the transformations $\varphi$ of the
synchronization group, treated as morphisms, form a category
 \begin{equation}
 {\cal A}=\left(M_w,\varphi\right). \label{cond7}
 \end{equation}
The action $\varphi$ of the synchronization group
$L_{+}^{\uparrow\,(S)}$ is defined in the most natural way
 \begin{equation}
 \varphi(M_w)=M_{\Lambda^S\circ w},\qquad
 \Lambda^S\in L_{+}^{\uparrow\,(S)}. \label{cond8}
 \end{equation}
From the physical point of view all choices of the element of the
category ${\cal A}$ are equivalent provided that 
the relativity principle holds. 
However, if we want to  introduce covariant canonical formalism for
a relativistic free particle on the classical level or to define a
proper position operator for such a particle on the quantum level we
have to give up the relativity principle;
a consistent description is possible
only if we use a fixed element of the category ${\cal A}$.
In this case also causal problems connected with breaking of Bell's
inequalities in QM disappear.
Summarizing, formulation of special relativity in terms of the
category ${\cal A}$ is equivalent to the standard one, however,
whenever 
the notion of localizability or absolute causality are incorporated,
the group of morphisms is 
broken i.e.\ a concrete time orientation is fixed. The advantage of
the use of ${\cal A}$ (family of CT synchronizations) in comparison
with EP scheme is, that in the 
former case we can consistently define the position operator while in
the later one it is impossible. Moreover this construction shows,
that some notions (like localizability) are simultaneously
compatible with quantum mechanics and Poincar{\'e} covariance only if
we resign with democracy between inertial frames, i.e.\ if a
privileged frame is distinguished.

\section{Canonical formalism in the CT synchronization}
\label{pcaban:sec:canonicalformalism}
For a relativistic free particle we postulate the following
action functional
 \begin{equation}
 S_{12}=-m\int\limits_{\lambda_1}^{\lambda_2}\sqrt{ds^2},
 \label{20}
 \end{equation}
where $ ds^2=g_{\mu\nu}(u)\frac{dx^\mu}{d\lambda}\frac{dx^\nu}{\lambda}
d\lambda^2$ and $\lambda$ is a trajectory parameter. We define the
fourvelocity in the standard way:
$ \omega^\mu=\frac{dx^\mu}{d\lambda}=\dot{x}^\mu$.
Then the velocity has the form:
$\vec{v}=\frac{d\vec{x}}{dx^0}=\frac{\vec{\omega}}{\omega^0}$.
Choosing the parameter $\lambda$ as the lenght of the
trajectory, $d\lambda=\sqrt{ds^2}$, we obtain the following
condition
 \begin{equation}
 \omega^2=g_{\mu\nu}(u)\omega^\mu(u)\omega^\nu(u)=1,
 \label{21}
 \end{equation}
which implies
 \begin{equation}
 \dot{\omega}^\mu=\ddot{x}^\mu=0.
 \label{22}
 \end{equation}
Using the formula (\ref{17}) we can derive from (\ref{20})
the Lagrangian
 \begin{equation}
 L=-m\sqrt{\left(1+u^0\vec{u}\vec{v}\right)^2-\left(\vec{v}\right)^2}.
 \label{23}
 \end{equation}
Now we can calculate the canonical momenta
 \begin{equation}
 \pi_i = \frac{\partial L}{\partial v^i} =
 \frac{{m} \left[v^i - u^i u^0 (1 + u^0 \vec{u}
 \vec{v})\right]}{\sqrt{(1 + u^0 \vec{u} \vec{v})^2-(\vec{v})^2}} =
 -m \omega_i,
 \label{24}
 \end{equation}
and the Hamiltonian
 \begin{equation}
 H = \pi_k v^k - L = 
 \frac{1}{u^0}\left(\vec{u}\nad{\pi} +
 \sqrt{\left(\vec{u}\nad{\pi}\!\right)^2+\left(\nad{\pi}\!\right)^2+m^2}
 \right)=
 {m} \omega_0,
 \label{25}
 \end{equation}
where $\nad{\pi}=(\pi_1,\pi_2,\pi_3)$.
So the covariant fourmomentum can be defined by
 \begin{equation}
 k_\mu=m\omega_\mu.
 \label{26}
 \end{equation}
It is easy to check, that $k_\mu$ fulfills the following
dispersion relation
 \begin{equation}
 k^2=g^{\mu\nu}(u)k_\mu k_\nu=m^2.
 \label{27}
 \end{equation}
Now we introduce the Poisson bracket
 \begin{equation}
 \{A, B\} = -\left({\delta^\mu}_\nu - \frac{k^\mu u_\nu}{u k}\right)
 \left(\frac{\partial A}{\partial x^\mu}
 \frac{\partial B}{\partial k_\nu}
 - \frac{\partial B}{\partial x^\mu}
 \frac{\partial A}{\partial k_\nu}\right),
 \label{30}
 \end{equation}
where all variables $x^\mu$, $k_\nu$ are treated as independent; in
particular $k_0$ is not {\em a priori} connected with $k_i$ {\em via}
the disperssion 
relation (\ref{27}).
The Poisson bracket defined by the above formula satisfies all
necessary conditions:
\begin{list}{---}{\setlength{\parsep}{0pt}
\setlength{\topsep}{0pt}
\setlength{\itemsep}{0pt}}
\item it is linear with respect to both factors, antisymmetric,
  satisfies the Leibniz rule and fulfills the Jacobi identity;
\item it is manifestly Poincar\'e covariant in the CT synchronization;
\item it is consistent with the constraint (\ref{27}),
  i.e.\ $\left\{k^2,k_\nu\right\}=\left\{k^2,x^\mu\right\}=0$;
  therefore there is no reason to introduce a Dirac bracket.
\item it is consistent with the Hamilton equations
  (\ref{28});
\end{list}
In particular from (\ref{30}) we get
 {\renewcommand{\arraystretch}{1.2}
 \begin{equation}
 \begin{array}{llcllll}
 \left\{x^\mu,x^\nu\right\}&=&0,&\rule{0.5cm}{0cm}&
 \left\{x^0,k_\mu\right\}&=&0,\\ 
 \left\{x^i,k_j\right\}&=&-\delta^{i}_{j},&&
 \left\{x^i,k_0\right\}&=&\frac{k^i}{k^0},\\
 \left\{k_\mu,k_\nu\right\}&=&0.&&
 \end{array}
 \end{equation}}
The Hamilton equations for a free particle have the desirable form
 \begin{equation}
 \frac{dx^i}{dt} = \frac{\partial H}{\partial \pi_i} 
 = \frac{k^i}{k^0} = v^i,
 \qquad\quad
 \frac{dk_i}{dt} = -\frac{\partial H}{\partial x^i} = 0,
 \label{28}
 \end{equation}
where $H$ is given by (\ref{25}). In general, the equation of motion
for an observable $\Omega(x^\mu,k_\nu)$ expressed by the Poisson
bracket (\ref{30}) is
 \begin{equation}
 \frac{d\Omega}{dt}= \frac{\partial\Omega}{\partial t}+
 \{\Omega,k_0\}. 
 \label{obserwabla}
 \end{equation}
In the above equation as well as in (\ref{30}) $k_0$ is treated as
an independent variable. Solution of 
(\ref{obserwabla}) can be subduced to
the constraint surface (\ref{27}).

\section{Quantum theory in the CT synchronization}
\label{pcaban:sec:quantumtheory}
The results of previous sections imply that in the CT synchronization we
have absolute causality so in this approach disappear all
causal problems connected with violation of Bell inequalities. Quantum
theory remains non-local but it is causal. In our approach we are
able, in analogy to the classical Poisson algebra described in section
\ref{pcaban:sec:canonicalformalism}, to introduce a Poincar\'e
covariant algebra of momentum and position operators satisfying all
fundamental physical requirements. This is done in the section
\ref{pcaban:sec:algebraofmomenta}. Properties of the introduced
position operator will be 
discussed in details in the sections \ref{pcaban:sec:localisedstates}
and \ref{pcaban:sec:invariantmeasure}. 
\subsection{Preliminaries}
In the CT synchronization the following point of view is the
most natural one: with each inertial observer $O_u$ we connect
his own Hilbert space $H_u$ (space of states).
The states vectors from $H_u$ are denoted by $u$:
$\ket{u,\dots}$. In other words we have a bundle of Hilbert spaces 
corresponding to the bundle of frames described in sec.\
\ref{bundle}. In such an interpretation
we have to distinguish carefuly active and passive transformations,
because in our approach active transformations are represented by
operators acting in one Hilbert space while passive ones by
operators acting beetwen different Hilbert spaces.
So, in particular, the Lorentz group transformations are considered as
passive ones.
Now, let $U(\Lambda)$ be an operator representing a Lorentz
group element $\Lambda$. We postulate the following, standard,
transformation law for a contravariant fourvector operator
 \begin{equation}
 U(\Lambda)\hat{A}(u)^\mu U^{-1}(\Lambda)=
 \left(D^{-1}(\Lambda,u)\right)^{\mu}_{\,\,\nu}\hat{A}(u^\prime)^\nu;
 \label{31}
 \end{equation}
where $D(\Lambda,u)$ is given by eqs.\ (\ref{10},\ref{11}) and
$u^\prime=D(\Lambda,u)u$; for a covariant fourvector $\hat{A}(u)_\mu$
we have to replace $D^{-1}$ by $D^T$  on the right hand side of
(\ref{31}). 
Let $\Omega$ be a fourvector observable.
In the space $H_u$ the observable $\Omega$ is represented by
an operator $\hat{\Omega}^\mu(u)$. Now let two inertial
observers $O_u$ and $O_{u^\prime}$ measure independently the value
of the observable $\Omega$ for a physical system being in
the same physical state\footnote{Of course, we should imagine an
ansamble of identical copies of a physical system in the
same prepared state.}.
Let this state in the space $H_u$ be described by the eigenvector
$\ket{\omega,u,\dots}$ of the $\Omega^\mu$. Then in the space
$H_{u^\prime}$ the same state is described by the vector
 \begin{equation}
 \left|\omega^\prime,u^\prime,\dots\right>=U(\Lambda)
 \left|\omega,u,\dots\right>,
 \label{32}
 \end{equation}
where $u^\prime=D(\Lambda,u)u$, $\omega^\prime=D(\Lambda,u)\omega$.
As a result of measurement of $\Omega$ the observer $O_u$ will receive
the value $\omega$
 \begin{equation}
 \hat{\Omega}^\mu(u)\left|\omega,u,\dots\right>=
 \omega^\mu \left|\omega,u,\dots\right>.
 \label{33}
 \end{equation}
As a result of measurment the observer $O_{u^\prime}$ should
obtain the value $\omega^\prime=D(\Lambda,u)\omega$.
So in the space of states $H_{u^\prime}$ the observable $\Omega$
is represented by an operator $\hat{\Omega}^\mu(u^\prime)$,
because
 \begin{eqnarray}
 \lefteqn{\hat{\Omega}(u^\prime)\left|\omega^\prime,u^\prime,\dots\right>
 =U(\Lambda)D(\Lambda,u)\hat{\Omega}(u)
 \left|\omega,u,\dots\right>=\mbox{}}\nonumber\\
 &&\mbox{}=U(\Lambda)D(\Lambda,u)\omega(u)
 \left|\omega,u,\dots\right>=\mbox{}\nonumber\\
 &&\mbox{}=\omega^\prime(u^\prime)
 \left|\omega^\prime,u^\prime,\dots\right>,
 \label{34}
 \end{eqnarray}
where we have used (\ref{31},\ref{32}).
To conclude this section we provide the interpretation of the operator
$\hat{\Omega}^\prime(u^\prime)=D(\Lambda,u)\hat{\Omega}(u)$.
We have
 \begin{eqnarray}
 \lefteqn{\hat{\Omega}^{\prime\mu}(u^\prime)\left|\omega,u,\dots\right>=
 D^{\mu}_{\ \nu}(\Lambda,u)\hat{\Omega}^{\nu}(u)
 \left|\omega,u,\dots\right>=\mbox{}}\nonumber\\
 &&\mbox{}=D^{\mu}_{\ \nu}(\Lambda,u)\omega^\nu(u)
 \left|\omega,u,\dots\right>=\omega^{\prime\mu}(u^\prime)
 \left|\omega,u,\dots\right>.
 \label{35}
 \end{eqnarray}
Thus $\hat{\Omega}^\prime(u^\prime)$ is an operator when acting on
a vector describing the state of a physical system in the space
of the observer $O_u$ gives the same result as seen by the
observer $O_{u^\prime}$ performing measurement on this system
being in the same physical state.

\subsection{Algebra of momenta and positions}
\label{pcaban:sec:algebraofmomenta}
We can introduce in each space $H_u$ the Hermitean fourmomentum
operators $\hat{p}_\lambda(u)$ (generators of translations).
These operators are interpreted as observables in the 
corresponding reference frame . In the CT synchronization we can go
further and introduce in each space $H_u$ Hermitean
position operators $\hat{x}^\mu(u)$. 
According to the Poisson bracket on the classical
level (\ref{30}) we postulate the following commutators between
$\hat{x}^\mu(u)$ and $\hat{p}_\lambda(u)$ 
  \begin{equation}
 [\hat{x}^\mu(u),\hat{p}_\lambda(u)]=i\left(
 \frac{u_\lambda\hat{p}^\mu(u)}{u\hat{p}(u)} - 
 \delta^{\mu}_{\ \lambda}\right),
 \label{50}
 \end{equation}
 \begin{equation}
 [\hat{p}_\mu(u),\hat{p}_\nu(u)]=0,
 \label{47}
 \end{equation}
 \begin{equation}
 [\hat{x}^\mu(u),\hat{x}^\nu(u)]=0.
 \label{54}
 \end{equation}
In particular
 \begin{equation}
 \left[\hat{x}^0(u),\hat{p}_\lambda(u)\right]  =  0,
 \label{51}
 \end{equation}
 \begin{equation}
 \left[\hat{x}^i(u),\hat{p}_j(u)\right]  =  -i\,\,\delta^{i}_{\ j},
 \label{52}
 \end{equation}
 \begin{equation}
 \left[\hat{x}^i(u),\hat{p}_0(u)\right]  = 
 i\,\,\frac{\hat{p}^i(u)}{\hat{p}^0(u)}.
 \label{53}
 \end{equation}
We see that $\hat{x}^0$ commutes with
all observables, it allows us to interprete $\hat{x}^0$ as a parameter
just like in the standard nonrelativistic quantum mechanics.
We have to stress here once again that above commutation
relations defining position operators {\em are covariant} in the
CT synchronization. It can be checked directly; one simply has
to use the eqs.\ (\ref{10},\ref{11},\ref{31}) and to transform the eqs.\
(\ref{50},\ref{47},\ref{54}) to another reference frame.
One can also check that
 \begin{equation}
 [\hat{x}^\mu(u),\hat{p}^2]=0,
 \label{55}
 \end{equation}
which means that localized states have definite mass.

\section{Unitary orbits of the Poincar{\'e} group for $k^2>0$ and spin
  operators}
According to our interpretation we deal with a bundle
of Hilbert spaces $H_u$ rather than with a single space of
states. Therefore transformations of the Lorentz group 
induce an orbit in this bundle. In this section we
construct and classify unitary orbits of the Poincar\'e group in the
above mentioned bundle of Hilbert spaces. As we will see, the unitary
orbits are classified with help of mass and spin, similarly as for the
standard unitary representations of the Poincar\'e group.
\subsection{Unitary orbits}
As in the standard case we assume that the eigenvectors
$\ket{k,u,\dots}$ of the fourmomentum operators
 \begin{equation}
 \hat{p}_\mu(u)\ket{k,u,\dots}=k_\mu\ket{k,u,\dots}
 \label{56}
 \end{equation}
with $k^2=m^2$, form a base in the Hilbert space $H_u$. We adopt the
following Lorentz-covariant normalization
 \begin{equation}
 \bracket{k^\prime,u,\dots}{k,u,\dots}=
 2\,k^0\,\delta^3(\nad{k}^\prime-\nad{k}\!),
 \label{57}
 \end{equation}
where $\nad{k}$ denotes the space part of the covariant
fourvector $k_\mu$ and $k^0=g^{0\mu}k_\mu$ is positive.
Energy $k_0$ is the solution of the dispersion relation
$k^2=m^2$ and is given by
 \begin{equation}
 k_0=\frac{1}{u^0}\left(-\vec{u}\nad{k} +
 \sqrt{\left(\vec{u}\nad{k}\!\right)^2+\left(\nad{k}\!\right)^2+m^2}
 \right),
 \label{58}
 \end{equation}
so
 \begin{equation}
 k^0\equiv\omega(\nad{k}\!)=u^0
 \sqrt{\left(\vec{u}\nad{k}\!\right)^2+\left(\nad{k}\!\right)^2+m^2}.
 \label{59}
 \end{equation}
In the construction of the unitary irreducible orbits we 
use the operator $e^{-i{\nad{\scriptstyle
 q}}\!\hat{\vec{x}}(u)}$. Action of this 
operator on the basis states can be determined by using its unitarity,
normalization of the basis vectors (\ref{57}) and the
commutation relations (\ref{50}). Its final form is\footnote{There is
a freedom in choosing the phase factor on the right hand side of
(\ref{60}); it is determined here by the requirement that no phase
factor on the right hand side of (\ref{68}) appears.}
 \begin{equation}
 e^{-i\,\nad{\scriptstyle q}\!\hat{\vec{x}}(u)}\ket{k,u,\dots}=
 e^{iq_0\hat{x}^0(u)}
 \sqrt{\frac{uk}{u(k+q)}}\ket{k+q,u,\dots},
 \label{60}
 \end{equation}
where on the right hand side of (\ref{60}) $q_0$ is determined by
$\nad{k}$, $\nad{q}$ and $u$; namely
 \begin{eqnarray}
 \lefteqn{q_0=\frac{1}{u^0}\left(-\vec{u}\nad{q}\! - 
 \sqrt{\left(\vec{u}\nad{k}\!\right)^2+\left(\nad{k}\!\right)^2+m^2}+
 \null\right.}\nonumber\\
 &&\left.\null+\sqrt{\left(\vec{u}\nad{q}\!+\vec{u}\nad{k}\!\right)^2+
 \left(\nad{k}\!+\nad{q}\!\right)^2+m^2}\right).
 \label{61}
 \end{eqnarray}
The basis vectors of the space $H_u$ can be generated from a vector
representing a particle at rest with respect to the preferred
frame. Firstly we act $U(L_u)$ on such a vector; the resulting state
has fourmomentum $mu_\mu$ and belongs to $H_u$. Next, by means of the
formula (\ref{60}), we generate in $H_u$ a vector with fourmomentum
$k_\mu$. Precisely
 \begin{equation}
 \ket{k,u,\dots}=\sqrt{\frac{uk}{m}}
 e^{-i\,(k_\mu-mu_\mu)\,\hat{x}^\mu(u)} 
 U(L_u)\ket{\Nad{k}\!,\tilde{u},\dots},
 \label{62}
 \end{equation}
where:
 \begin{equation}
 \tilde{u}=(1,\vec{0}),\qquad \Nad{k}\!=(m,\nad{0}\!),\qquad
 u=D(L_u,\tilde{u})\tilde{u}.
 \label{64}
 \end{equation}
The above mentioned orbit induced by the action of the operator
$U(\Lambda)$ in the bundle of Hilbert spaces is fixed
by the following covariant conditions
\begin{list}{---}{\setlength{\parsep}{0pt}
\setlength{\topsep}{0pt}
\setlength{\itemsep}{0pt}}
\item $k^2=m^2$;
\item $\varepsilon(k^0)=inv.$, for physical representations
 $k^0>0,\,\varepsilon(k^0)=1$.
\end{list}
As a consequence there exists a positive defined, Lorentz
invariant measure 
 \begin{equation}
 d\mu(k,m)=d^4k\,\theta(k^0)\, \delta(k^2-m^2).
 \label{67}
 \end{equation}
Now, applying the Wigner method and using eq.\ (\ref{31})
one can easily determine the action of
the operator $U(\Lambda)$ on the basis vector. We find
 \begin{equation}
 U(\Lambda)\ket{k,u,m;s,\sigma}=
 {{\cal D}_{\sigma\lambda}^{s}}^{-1}(R_{\Lambda,u})
 \ket{k^\prime,u^\prime,m;s,\lambda},
 \label{68}
 \end{equation}
where
 \begin{equation}
 u^\prime=D(\Lambda,u)u=D(L_{u^\prime},\tilde{u})\tilde{u},
 \label{69}
 \end{equation}
 \begin{equation}
 k^\prime=D^{T-1}(\Lambda,u)k,
 \label{70}
 \end{equation}
 \begin{equation}
 R_{\Lambda,u}=D(R_{\Lambda,u},\tilde{u})=
 D^{-1}(L_{u^\prime},\tilde{u})
 D(\Lambda,u) D(L_u,\tilde{u})\subset SO(3)
 \label{71}
 \end{equation}
and ${\cal D}^{s}_{\sigma\lambda}(R_{\Lambda,u})$ is the standard spin
$s$ rotation matrix $s=0,\frac{1}{2},1,\dots$;
$\sigma,\lambda=-s,-s+1,\dots,s-1,s$. 
$D(R_{\Lambda,u},\tilde{u})$ is the Wigner rotation belonging
to the little group of a vector $\tilde{u}$. Let us stress that in our
approach, 
contrary to the standard one, representations of the Poincar\'e
group are induced from the little group of a vector $\tilde{u}$,
not $\Nad{k}$.
Finally, the normalization (\ref{57}) takes the form
 \begin{equation}
 \bracket{k,u,m;s,\lambda}
 {k^\prime,u,m;s^\prime,\lambda^\prime}=
 2k^0\delta^3(\nad{k}\!^\prime-\nad{k}\!)\delta_{s^\prime s}
 \delta_{\lambda^\prime\lambda}.
 \label{72}
 \end{equation}

\subsection{Spin}
Now we describe in some details transformation properties of
a second rank covariant tensor operator. These results
are used in the discussion of the spin.\\
Let $\hat{M}(u)=[\hat{M}_{\mu\nu}(u)]$ be a tensor operator. The
transformation law for this tensor can be deduced from (\ref{31}) and
can be written in the matrix notation as
 \begin{equation}
 U(\Lambda)\hat{M}(u)U^{-1}(\Lambda)=
 D^T(\Lambda,u)\hat{M}(u^\prime)D(\Lambda,u). 
 \label{Spin1}
 \end{equation}
The lower--triangular form of the matrix $D(\Lambda,u)$
(see eq.\ (\ref{10}) and (\ref{11})) 
implies that the space part of $\hat{M}$ transform into itself, 
namely
 \begin{equation}
 U(\Lambda)\hat{M}_{ij}(u)U^{-1}(\Lambda)=
 \Omega_{ki}(\Lambda,u)\hat{M}_{kl}(u^\prime)\Omega_{lj}(\Lambda,u), 
 \label{Spin2}
 \end{equation}
where $\Omega(\Lambda,u)$ denotes the space part of the matrix
$D(\Lambda,u)$. 
By means of the triangular form of $D(\Lambda,u)$ it is easy to see
that
 \begin{equation}
 g_{ij}(u^\prime)=\Omega_{ki}^{-1}(\Lambda,u)g_{kl}(u)
 \Omega_{lj}^{-1}(\Lambda,u),
 \label{Spin3}
 \end{equation}
where $g_{ij}$ are the space components of the covariant metric tensor
$g_{\mu\nu}$.
Therefore, denoting by $\gamma_{ij}$
the inverse of the matrix of the space part of the
$g_{\mu\nu}(u)$, that is
 \begin{equation}
 \gamma_{ij}(u)=[g_{ij}]^{-1}=-(\delta_{ij}+u^iu^j)
 \label{Spin4}
 \end{equation}
one can easily show that the bilinear form
 \begin{equation}
 \hat{M}^2=\gamma_{ik}(u)\gamma_{jl}(u)\hat{M}_{ij}(u) \hat{M}_{kl}(u)
 \label{Spin5}
 \end{equation}
is a Poincar\'e invariant operator.
Let us introduce the spin operators $\hat{S}_{ij}(u)$ transforming
covariantly according to (\ref{Spin2}) such that
 \begin{equation}
 U(\Lambda)\hat{S}_{ij}(u)U^{-1}(\Lambda)=
 \Omega_{ki}(\Lambda,u)\hat{S}_{kl}(u^\prime)\Omega_{lj}(\Lambda,u), 
 \label{Snowe1}
 \end{equation}
and defined by the action on the basis vectors
 \begin{equation}
 \hat{S}_{ij}(u)\ket{k,u,m;s,\lambda}=-{\cal S}_{ij}^{s}(u)_{\lambda\sigma}
 \ket{k,u,m;s,\sigma}. 
 \label{Snowe2}
 \end{equation}
By means of the equation (\ref{68}), the eqs.\ (\ref{Snowe1}) and
(\ref{Snowe2}) imply the following consistency condition
 \begin{equation}
 {\cal D}^{s}(R_{\Lambda,u}){\cal S}_{ij}^{s}(u)
 {\cal D}^{s}(R^{-1}_{\Lambda,u})=\Omega_{ki}(\Lambda,u)
 {\cal S}^{s}_{kl}(u^\prime)\Omega_{lj}(\Lambda,u).
 \label{Snowe3}
 \end{equation}
Therefore, using the fact that $R_{L_u,\tilde{u}}=I$ and 
 \begin{equation}
 D(L_u,\widetilde{u})=
 \left(\begin{array}{c|c}
 u^0 & 0 \\
 \hline
 \vec{u} & I + \frac{u^0}{1+u^0}\vec{u}\otimes\vec{u}^T
 \end{array}\right)
 \label{600}
 \end{equation}
(so $\Omega_{ij}(L_u,\tilde{u})=\delta_{ij}+\frac{u^0}{1+u^0}u^iu^j$),
one obtains
 \begin{equation}
 {\cal S}^{s}_{ij}(u)=\tilde{\cal S}^{s}_{ij}+
 \frac{(u^0)^2}{1+u^0}(u^j\delta_{li}-u^i\delta_{lj})
 u^k\tilde{\cal S}^{s}_{kl},
 \label{Snowe4}
 \end{equation}
where $\tilde{\cal S}^{s}_{ij}:={\cal S}^{s}_{ij}(\tilde{u})$ are
assumed to be Hermitean matrix generators of the unitary
representation ${\cal D}^s(R)$ of $SO(3)$, i.e.\
 \begin{equation}
 \tilde{\cal S}^{s}_{ij}=-\tilde{\cal S}^{s}_{ji}=
 \left(\tilde{\cal S}^{s}_{ij}\right)^\dagger,
 \label{Snowe5}
 \end{equation}
and
 \begin{equation}
 \left[\tilde{\cal S}^{s}_{ij},\tilde{\cal S}^{s}_{kl}\right]=
 -i\left(\delta_{il}\tilde{\cal S}^{s}_{jk}+
 \delta_{jk}\tilde{\cal S}^{s}_{il}-
 \delta_{ik}\tilde{\cal S}^{s}_{jl}-
 \delta_{jl}\tilde{\cal S}^{s}_{ik}\right).
 \label{Snowe6}
 \end{equation}
Consequently, in an arbitrary frame
 \begin{equation}
 {\cal S}^{s}_{ij}(u)=-{\cal S}^{s}_{ji}(u)=
 \left.{\cal S}^{s}_{ij}\right.^\dagger(u),
 \label{Snowe7}
 \end{equation}
and
 \begin{eqnarray}
 \lefteqn{\left[{\cal S}^{s}_{ij}(u),{\cal
 S}^{s}_{kl}(u)\right]=\mbox{}}
 \label{Snowe8} \\
 &&\mbox{}=i\left(g_{il}(u){\cal S}^{s}_{jk}(u)+
 g_{jk}(u){\cal S}^{s}_{il}(u)-
 g_{ik}(u){\cal S}^{s}_{jl}(u)-
 g_{jl}(u){\cal S}^{s}_{ik}(u)\right).\nonumber
 \end{eqnarray}
Therefore, the spin operators $\hat{S}_{ij}(u)=-\hat{S}_{ji}(u)$ are
Hermitean and satisfy the same algebra
 \begin{eqnarray}
 \lefteqn{\left[{\hat S}_{ij}(u),{\hat S}_{kl}(u)\right]=\mbox{}}
 \label{Snowe9} \\
 &&\mbox{}=i\left(g_{il}(u){\hat S}_{jk}(u)+
 g_{jk}(u){\hat S}_{il}(u)-
 g_{ik}(u){\hat S}_{jl}(u)-
 g_{jl}(u){\hat S}_{ik}(u)\right).\nonumber
 \end{eqnarray}
Now, according to (\ref{Spin5}) one can define the invariant spin
square operator
 \begin{eqnarray}
 \hat{S}^2 & = & \frac{1}{2}\gamma_{ik}(u)\gamma_{jl}(u)\hat{S}_{ij}(u)
 \hat{S}_{kl}(u)=\mbox{} \label{Snowe10}\\
 \mbox{} & = & \frac{1}{2}\hat{S}_{ij}(u)\hat{S}_{ij}(u)+
 u^iu^j\hat{S}_{ik}(u)\hat{S}_{jk}(u).\nonumber
 \end{eqnarray}
Consequently
 \begin{equation}
 \hat{S}^2\ket{k,u,m;s,\lambda}=s(s+1)\ket{k,u,m;s,\lambda}.
 \label{Snowe11}
 \end{equation}
Finally, as follows from (\ref{Snowe2}), (\ref{56}) and (\ref{60}),
$\hat{S}_{ij}(u)$ commute with $\hat{p}_\mu(u)$ and $\hat{x}^\mu(u)$
i.e.
 \begin{equation}
 \left[\hat{S}_{ij}(u),\hat{p}_\mu(u)\right]=
 \left[\hat{S}_{ij}(u),\hat{x}^\mu(u)\right]=0.
 \label{Snowe12}
 \end{equation}
Summarizing, the introduced covariant spin operator has properties
showing its advantage in comparison with the standard one. In
particular the algebra generated by $\hat{p}_\mu(u)$, $\hat{x}^\mu(u)$
and $\hat{S}_{ij}(u)$ -- the equations
(\ref{50},\ref{47},\ref{54},\ref{Snowe9},\ref{Snowe12}) --
is evidently covariant under the Poincar\'e group action.

\section{The position operator and localized states}
In the section \ref{sec:pcaban:NewtonWigner} we recall briefly the
Newton--Wigner position operator. The section
\ref{pcaban:sec:localisedstates} is devoted to localized states and
derivation of a functional 
form of the introduced in the section \ref{pcaban:sec:quantumtheory}
position operator. Section  
\ref{pcaban:sec:invariantmeasure} is devoted to description of
localized states and position operator in the Hilbert space with a
fully invariant measure, resembling the nonrelativistic one.
\subsection{The Newton--Wigner operator}\label{sec:pcaban:NewtonWigner}
In the non-relativistic quantum mechanics the situation is clear,
we can define the position operator which fulfills all  the
conditions stated in the introduction (covariance is of course
understood with respect to the Galilei group). Its construction
and properties are very well known and we do not intent to describe
them in this section.
In the relativistic case situation is much more complicated.
One of the earliest definitions of the position operator is due
to Newton and Wigner \cite{NewWig}. In this approach the authors
try first to find states of the particle localized at a given
point $(t,\vec{a})$ and then to write down the corresponding
position operators.
Let $S_{\vec{a}}$, the set of states $\psi_{\vec{a},0}$ localized
at $\vec{a}\in {\mathbb{R}}^3$ at $t=0$, be the subset of the
Hilbert space $\cal{H}$ of the unitary irreducible representation
of the universal covering group of the Poincar\'e group.
The Newton--Wigner postulates are as follows:
\begin{list}{---}{\setlength{\parsep}{0pt}
\setlength{\topsep}{0pt}
\setlength{\itemsep}{0pt}}
\item the set $S_{\vec{a}}$ is a linear subspace of $\cal{H}$;
\item $S_{\vec{a}}$ is invariant under rotations around point
  $\vec{a}$, reflections in $\vec{a}$, and time inversions;
\item $S_{\vec{a}}$ is orthogonal to all its space translates,
  i.e.\ under the space translations each
  $\psi_{\vec{a},0}\in S_{\vec{a}}$ transforms to a state
  from $\cal{H}$ which is orthogonal to all states from
  $S_{\vec{a}}$;
\item certain regularity conditions.
\end{list}
As an example let us discuss shortly the Newton--Wigner position
operator for a spinless particle. In this case $\cal{H}$ is a linear
space of solutions to the Klein--Gordon equation with positive
energy. Using the Fourier transform one can obtain the states
localized at $\vec{a}\in{\mathbb{R}}^3$ at $t=0$ in the momentum
representation, namely
 \begin{equation}
 S_{\vec{a}}=\left\{\psi_{\vec{a},0}(k)=
 \frac{1}{(2\pi)^{3/2}}k_{0}^{1/2}e^{-i\vec{k}\vec{a}}\right\}.
 \label{1}
 \end{equation}
The corresponding position operators are given by:
 \begin{equation}
 \hat{q}^k=-i\left(\frac{\partial}{\partial k_i} +
 \frac{1}{2}\frac{k^i}{(\vec{k})^2+m^2}\right).
 \label{2}
 \end{equation}
The main results obtained by Newton and Wigner may be summarized
in the following way
\begin{list}{}{\setlength{\parsep}{0pt}
\setlength{\topsep}{0pt}
\setlength{\itemsep}{0pt}}
\item[---] position operator exists for massive particles
  with arbitrary spin,
\item[---] it is hermitean,
\item[---] it has commuting components,
\item[---] under rotations transforms like a vector,
\item[---] $[\hat{q}^k,\hat{k}_i]=-i\delta_{i}^{k}$
   (canonical commutation
   relations with momentum operators),
\item[$\bullet$] it is not covariant,
\item[$\bullet$] localized states are not covariant,
\item[$\bullet$] massles particles with spin are not localizable.   
\end{list}
Of course a lot of trials have been undertaken to remove all the
unsatisfactory features of the Newton--Wigner approach, but to
the best knowledge of the authors no one of them has been fully
succesfull. For a review see \cite{Bac,Nied}.

\subsection{Localized states and momentum representation of the
  position operator}
\label{pcaban:sec:localisedstates}
In this section we briefly describe some properties of
the introduced in the section \ref{pcaban:sec:algebraofmomenta} 
position operator. Firstly let us find localized states
in the Schr\"odinger picture. 
Taking into account the eqs.\ (\ref{55}) and (\ref{Snowe12})
we find that $\hat{x}^\mu(u)$ 
commutes with $\hat{p}^2$ and $\hat{S}_{ij}(u)$ so all these three
operators have common eigenvectors and consequently localized
states have definite mass and spin. Let
$\ket{\vec{\xi},\tau,u,m;s,\lambda}$ 
denotes a state localized at the time $\tau$ in the space point
$\vec{\xi}$
 \begin{equation}
 \hat{\vec{x}}(u)\ket{\vec{\xi},\tau,u,m;s,\lambda}=
 \vec{\xi}\ket{\xi,\tau,u,m;s,\lambda}. 
 \label{119} 
 \end{equation}
The state $\ket{\vec{\xi},\tau,u,m;s,\lambda}$ can be expressed with
help of the invariant measure (\ref{67}) in
terms of the basis vectors $\ket{k,u,m;s,\lambda}$, namely
 \begin{equation}
 \ket{\vec{\xi},\tau,u,m;s,\lambda}=\frac{1}{(2\pi)^{3/2}} \int
 \frac{d^3\nad{k}}{2\omega(\nad{k}\!)} \sqrt{uk}\,
 e^{i\nad{k}\!\vec{\xi}}  \ket{k,u,m;s,\lambda}.
 \label{123}
 \end{equation}
Now, after an arbitrary time $t$ this state evolves to
 \begin{equation}
 \ket{\vec{\xi},\tau,u,m;s,\lambda;t}=\frac{1}{(2\pi)^{3/2}}
 \int\frac{d^3k}{2\omega(\nad{k})}\sqrt{uk} e^{ik_\mu\xi^\mu} 
 \ket{k,u,m;s,\lambda}
 \label{500}
 \end{equation}
with $\xi^0=\tau-t$.
One can easily check that these states are normalized as follows
 \begin{equation}
 \bracket{\vec{\xi}^\prime,\tau,u,m;s^\prime,\lambda^\prime;t}
 {\vec{\xi},\tau,u,m;s,\lambda;t}= 
 \frac{1}{2u^0}\delta^3(\vec{\xi}-\vec{\xi}^\prime)
 \delta_{ss^\prime} \delta_{\lambda\lambda^\prime}.
 \label{124}
 \end{equation}
It is worthwhile to notice here that the states given by
the eq.\ (\ref{123}) are covariant in the CT synchronization, i.e.\ a
state localized in the time $t=\tau$ for the observer $O_u$ is
localized in the time
$t^\prime=\tau^\prime=D^{0}_{\,\,0}(\Lambda,u)\tau$ for the observer
$O_{u\prime}$ too. 
Let us discuss a realization of the position operator in the
momentum representation.
Wave functions in momentum representation are
defined in the standard way
 \begin{equation}
 \psi^{m,s}_{\lambda}(k,u)=\bracket{k,u,m;s,\lambda}{\psi},
 \label{125}
 \end{equation}
or equivalently
 \begin{equation}
 \ket{\psi}=\sum\limits_{\lambda}\int\frac{d^3\nad{k}}{2\omega(\nad{k}\!)}
 \psi^{m,s}_{\lambda}\ket{k,u,m;s,\lambda}.
 \label{126}
 \end{equation}
The scalar product is given by
 \begin{equation}
 \bracket{\varphi}{\psi}= \sum\limits_{\lambda}\int
 \frac{d^3\nad{k}}{2\omega(\nad{k}\!)}
 \varphi^{*m,s}_{\phantom{*}\lambda}(k,u)
 \psi^{m,s}_{\lambda}(k,u).
 \label{127}
 \end{equation}
Now we can identify the wave functions related to the localized states
(\ref{123}); namely, we have
 \begin{equation}
 \chi^{m,s}_{\lambda}(\vec{\xi},\tau,k,u;\sigma;t)=
 \frac{1}{(2\pi)^{3/2}}\sqrt{uk}\, 
 e^{ik_\mu\xi^\mu}\delta_{\sigma\lambda}.
 \label{128}
 \end{equation}
It follows that in this realisation
 \begin{equation}
 \hat{x}^i=-i\frac{\partial}{\partial k_i} + \frac{1}{2}i
 \left(\frac{u^i}{uk}-\frac{k^i}{(uk)^2}\right).
 \label{130}
 \end{equation}
Evidently, for $\xi^0=0$ (i.e.\ for $t=\tau$) the functions $\chi$ are
eigenvectors of $\hat{x}^i$.
It can be easily demonstrated that in the preferred frame
($u=(1,\vec{0})$) the
function (\ref{128}) reduces to the Newton--Wigner localized state
(\ref{1}); then also the operator (\ref{130}) coincides 
with the Newton--Wigner position operator (\ref{2}) for a spinless
particle. 

\subsection{Invariant measure}
\label{pcaban:sec:invariantmeasure}
In the previous sections we used the Lorentz invariant
measure (\ref{67})
 \[
 d\mu(k,m)=d^4k\,\theta(k^0)\, \delta(k^2-m^2).
 \]
We point out that this measure is not invariant under the action of
the operator 
$e^{-i\nad{\scriptstyle q}\!\hat{\vec{x}}(u)}$ (compare eq.\ (\ref{60})). 
Nevertheless it is possible
to find the measure which is both Poincar\'e invariant and invariant
under the action of this operator.
One can easily check that such a measure can be written as
 \begin{equation}
 d\overline{\mu}(k,m)=uk\,d\mu(k,m)=uk\,d^4k\,\delta(k^2-m^2)
 \theta(k^0).
 \label{131}
 \end{equation}
Our motivation to introduce the measure (\ref{131}) is that it
simplifies
some of the formulas discussed herein and resembles the
nonrelativistic one. Let us integrate
the measure $d\overline{\mu}(k,m)$ with respect to the $k_0$.
We find
 \begin{equation}
 \int d\overline{\mu}(k,m)f(k)=
 \frac{1}{2u^0}\int d^3\nad{k} f(k_{0},\nad{k}),
 \label{132}
 \end{equation}
where $k_{0}$ is given by (\ref{58}).
Now the normalization (\ref{72}) is not
invariant under the action of the operator
$e^{-i\nad{\scriptstyle q}\hat{\vec{x}}(u)}$. 
To make it invariant we introduce rescaled basis vectors
 \begin{equation}
 \ket{k,u,m;s,\lambda}_{inv}:=
 \frac{1}{\sqrt{uk}}\ket{k,u,m;s,\lambda}.
 \label{134}
 \end{equation}
The rescaled vectors are normalized as follows:
 \begin{equation}
 \null_{inv}\bracket{k,u,m;s,\lambda}%
 {k^\prime,u,m;s^\prime,\lambda^\prime}_{inv}=
 2u^0\delta^3(\nad{k}-\nad{k}^\prime) \delta_{ss^\prime}
 \delta_{\lambda\lambda^\prime}.
 \label{135}
 \end{equation}
This normalization is invariant under the action of the operator
$e^{-i\nad{\scriptstyle q}\!\hat{\vec{x}}(u)}$ and it is
simultaneously Lorentz invariant. Moreover
 \begin{equation}
 e^{-i\,\nad{\scriptstyle q}
 \hat{\vec{x}}(u)}\ket{k,u,m;s,\lambda}_{inv}= 
 e^{iq_0t}\ket{k+q,u,m;s,\lambda}_{inv},
 \label{136}
 \end{equation}
where $q_0=q_0(\nad{k},\nad{q},u)$ is given by (\ref{61}).
The action of the operator $U(\Lambda)$ on the rescaled basis
vectors has again the form (\ref{68}), i.e.\
 \begin{equation}
 U(\Lambda)\ket{k,u,m;s,\sigma}_{inv}=
 {{\cal D}_{\sigma\lambda}^{s}}^{-1}(R_{\Lambda,u})
 \ket{k^\prime,u^\prime,m;s,\lambda}_{inv}.
 \label{137}
 \end{equation}
Now let us return to the position operator and localized states.
The localized states can be expressed in the new basis as follows
 \begin{equation}
 \ket{\vec{\xi},\tau,u,m;s,\lambda;t}=\frac{1}{(2\pi)^{3/2}} 
 \frac{1}{2u^0} \int d^3\nad{k} e^{ik_\mu\xi^\mu}
 \ket{k,u,m;s,\lambda}_{inv},
 \label{138}
 \end{equation}
where $\xi^0=\tau-t$.
The corresponding wave functions localized at the time $t=\tau$ take
the form 
 \begin{equation}
 \tilde{\chi}^{m,s}_{\lambda}(\vec{\xi},\tau,k,u;\sigma;t)=
 \frac{1}{(2\pi)^{3/2}} 
 e^{ik_\mu\xi^\mu}\delta_{\sigma\lambda},
 \label{141}
 \end{equation}
and the corresponding position operator takes the extremely simple
form 
 \begin{equation}
 \hat{x}^i=-i\frac{\partial}{\partial k_i}
 \label{142}
 \end{equation}
like in the nonrelativistic case.

\section{Conclusions}
According to suggestions of some authors 
(Bell \cite{Bell81}, Eberhard \cite{Eberhard78}), that a
consistent formulation of quantum mechanics demands existence of a
preferred frame, we constructed here a quantum
mechanics which has built--in the preferred frame and which is at the
same time Poincar\'e covariant.
We used introduced in \cite{Remb1} a nonstandard realisation of
Poincar\'e 
group; in this formulation the boost matrix has the lower-triangular
form so the time coordinate rescales only under Lorentz
transformations. Such a realisation corresponds to a nonstandard
synchronization of clocks (CT synchronization) i.e.\ to a different
than standard coordinate time definition. Clasically such a scheme is
operationally indistinguishable from the standard one.
Our construction shows,
that some notions (like causality, localizability) are simultaneously
compatible with quantum mechanics and Poincar{\'e} covariance only if
we resign with democracy between inertial frames, i.e.\ if a
privileged frame is distinguished. In this formulation of QM
causal problems connected with violation of Bell's
inequalities disappear. Quantum
theory remains in such a framework non-local but it is causal.
In this context we
constructed and classified unitary orbits of the Poincar\'e group in
the appriopriate bundle of Hilbert spaces. The unitary 
orbits are classified with help of mass and spin, similary as for the 
standard unitary representations of the Poincar\'e group, howeover are
induced differently from $SO(3)$.
We introduced a Poincar\'e
covariant algebra of momentum and position operators satisfying all
fundamental physical requirements. We proved that in our
framework the position operator 
exists for particles with arbitrary spin. It fulfills all the
requirements: it is Hermitean and covariant, it has commuting
components and moreover its eigenvectors (localised states) are
also covariant. We found the explicit functional form of the
position operator and demonstrated that
in the preferred frame our operator coincides with the
Newton--Wigner one. We also defined a covariant spin operators
and constructed an invariant spin square operator. Moreover, full
algebra of observables consisting of position operators, fourmomentum
operators and spin operators is manifestly Poincar\'e covariant in
this framework. 
We hope that this formulation may be usefull in the
construction of the Poincar\'e covariant version of the de
Broglie--Bohm quantum mechanics as well.

\section*{Acknowledgements}
This work was supported by the University of {\L}\'od\'z grant no.\ 621.

\end{document}